\documentclass[12pt]{article}
\usepackage{graphicx}

\newcommand\pubdate{\today}

\def\Title#1{\begin{center} {\Large #1 } \end{center}}

\def\Address#1{\begin{center}{ \it #1} \end{center}}

\newcommand\pubblock{\rightline{\begin{tabular}{l}  \\ 
         \pubdate  \end{tabular}}}
\newenvironment{Abstract}{\begin{quotation}  }{\end{quotation}}
\newenvironment{Presented}{\begin{quotation} \begin{center} 
             PRESENTED AT\end{center}\bigskip 
      \begin{center}\begin{large}}{\end{large}\end{center} \end{quotation}}

\begin{document}
\begin{titlepage}
 \pubblock
\vfill
\Title{Precision Measurement of the Longitudinal Double-Spin Asymmetry for Dijet Production at Intermediate Pseudorapidity in Polarized Proton+Proton Collisions at $\sqrt{s}$ = 200 GeV}
\vfill
	\begin{center}
		Zilong Chang$^{a}$ (speaker) and Ting Lin$^{b}$ for the STAR Collaboration
	\end{center}
\Address{$^{a}$CEEM, Indiana University - Bloomington, IN, USA}
\Address{$^{b}$Shandong University, China}
\vfill
\begin{Abstract}
Measurements of the longitudinal double-spin asymmetry, $A_{LL}$, by the STAR experiment have contributed significantly to our understanding of the gluon helicity distribution, $\Delta g(x)$, inside the proton. Results from the 2009 inclusive jet measurement, when included into global analyses, indicated a substantially positive polarization for gluons with partonic momentum fraction $x$ greater than 0.05. In addition to the inclusive jets, analyses of dijet production extending to higher pseudorapidity (up to $\eta \sim 1.8$) provide better constraints on the $x$ dependent behavior of $\Delta g(x)$. Recently, STAR published several new results at midrapidity (up to $\eta \sim 1.0$) using the $p+p$ data collected in 2012, 2013 and 2015 at both $\sqrt{s}$ = 510 and 200 GeV. These new results confirm the previous findings and provide additional constraints in the largely unexplored region of $x < 0.05$. In this talk, the preliminary results of the $A_{LL}$ for dijet production at intermediate pseudorapidity (up to $\eta \sim 1.8$) based on 2015 data at $\sqrt{s}$ = 200 GeV, with twice the figure-of-merit of the 2009 data, will be presented. The comparison with the theoretical expectations as well as its potential impact on $\Delta g(x)$ will be discussed.                                             
\end{Abstract}
\vfill
\newpage
\begin{Presented}
DIS2023: XXX International Workshop on Deep-Inelastic Scattering and
Related Subjects, \\
Michigan State University, USA, 27-31 March 2023 \\
     \includegraphics[width=9cm]{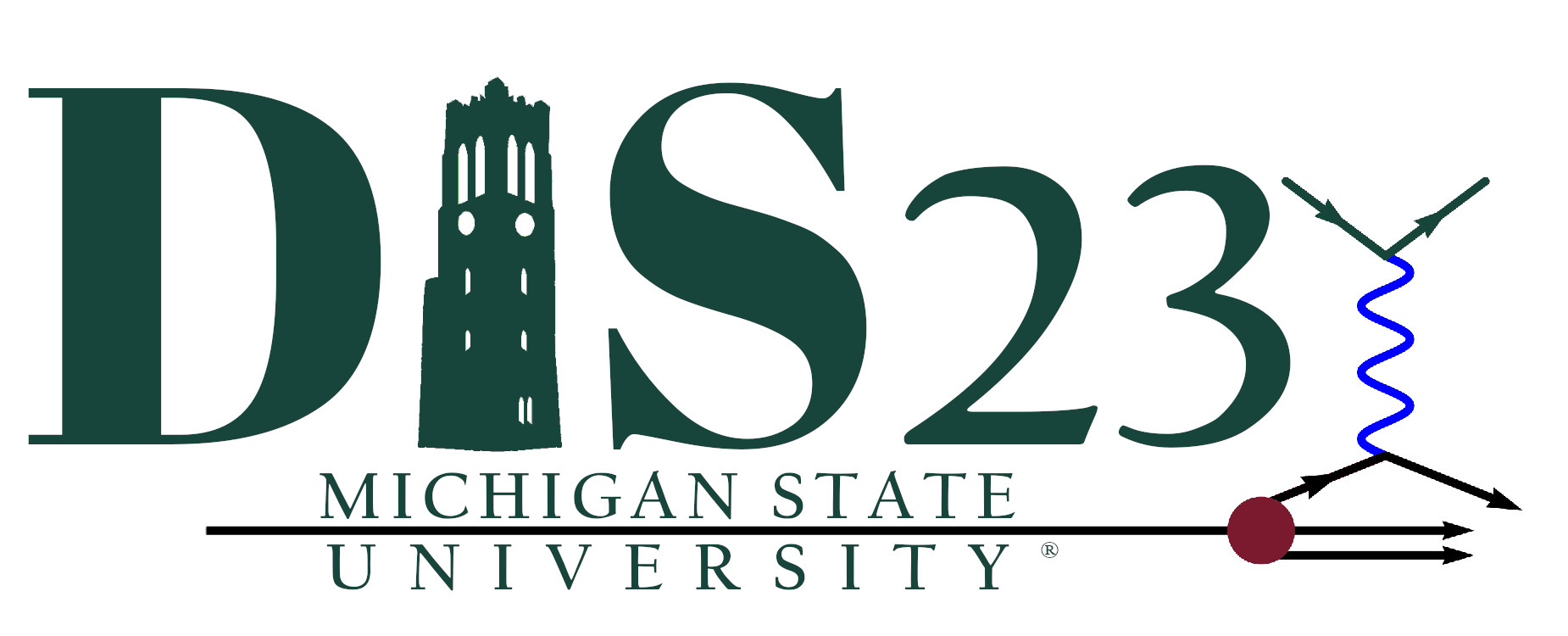}
\end{Presented}
\vfill
\end{titlepage}

\section{Introduction}
Exploring the spin structure and dynamics of the proton has been a major focus of the RHIC Spin program. Polarized DIS experiments have revealed that the quarks' spins account for only about 30\% of the proton's total spin, with the remainder attributed to contributions from the gluon spin and the orbital angular momenta of the partons~\cite{HERMES2007, COMPASS2017}. Understanding the helicity distribution of gluons within the proton is crucial to unraveling the internal structure and QCD dynamics of nucleons. The Relativistic Heavy Ion Collider (RHIC) at Brookhaven National Laboratory presents a unique opportunity to investigate gluon polarization through the collision of longitudinally polarized proton beams at center-of-mass energies of 200 and 510 GeV. At these energies, RHIC kinematics are highly sensitive to gluons, as the dominant scattering processes occur via the quark-gluon and gluon-gluon interactions.\par

High precision measurement of the longitudinal double spin asymmetry from STAR revealed for the first time, that there is a sizable polarization of the gluon inside the proton at momentum fractions ranging from $0.05 < x < 0.4$ \cite{deFlorian:2014yva, Nocera:2014gqa, Adamczyk:2015}. Extending the measurement to include dijet production allows for better constraints on the $x$ dependence of the gluon's helicity distribution. With more forward pseudorapidity measurements, our results can reach momentum fractions as low as $0.01$ with unprecedented precision than the previous experiments, thus providing more significant constraints to the largely unexplored low $x$ values of gluon's polarization inside the proton \cite{Adamczyk:2016okk, PhysRevD.98.032011, PhysRevD.98.032013, PhysRevD.100.052005, PhysRevD.103.L091103, PhysRevD.105.092011}.\par

\section{Gluon Polarization at RHIC}
The longitudinal double spin asymmetry, $A_{LL}$, is the observable used to explore the gluon polarization in this analysis. $A_{LL}$ is defined as the ratio of the longitudinally polarized cross section over the unpolarized one, and is roughly equal to
\begin{equation}
A_{LL} \sim \frac{\Delta f_{a}\Delta f_{b}}{f_{a}f_{b}}\hat{a}_{LL}
\end{equation}
where the $\Delta f_{a,b}$ are the helicity distributions of the two interacting partons. The $f_{a,b}$ are the unpolarized parton distribution functions, which have been well constrained by other experiments. $\hat{a}_{LL}$ is the partonic asymmetry that can be calculated from perturbative QCD and is very large at leading order.\par

RHIC is the world-only polarized hadron collider that is able to collide the longitudinally and transversely polarized protons at both 200 and 510 GeV~\cite{Alekseev:2003sk} . The Solenoidal Tracker at RHIC (STAR) \cite{Ackermann:2002ad} is a multi-purpose detector with several different sub-systems. The main sub-systems that are used in this analysis are the Time Projection Chamber (TPC)~\cite{Anderson:2003ur} and Electromagnetic Calorimeters~\cite{Beddo:2002zx, Allgower:2002zy}. TPC provides the measurement of the charged particle's momentum, charge discrimination as well as the particle identification with full azimuthal coverage over $|\eta| < 1.3$. The electromagnetic calorimetry is accomplished by Barrel Electromagnetic Calorimeter (BEMC) and Endcap Electromagnetic Calorimeter (EEMC). Both of them have full coverage in azimuthal angle, while BEMC covers the pseudorapidity from $-1 \le \eta \le 1$ and EEMC extends to $1 \le \eta \le 2$.\par

The data used in this analysis were recorded by STAR experiment during the 2015 running period with an integrated luminosity of 52 $\mathrm{pb^{-1}}$ and average polarization $58\%$ from longitudinally polarized proton+proton collisions at $\sqrt{s} = 200$ GeV. It is about twice the figure-of-merit of the previous published results using the 2009 data with an integrated luminosity of 21 $\mathrm{pb^{-1}}$. The same jet reconstruction and selection procedures were used; the anti-$k_{T}$ algorithm \cite{Cacciari:2008gp, Cacciari:2011ma} was used with resolution parameter R = 0.6. Multilayer perceptron, a supervised machine-learning regression provided by the ROOT TMVA library~\cite{ROOT:1997}, were used in order to correct for the sizable loss of charged particles in the endcap region due to the inefficiency of the tracking at $|\eta| > 1.3$~\cite{PhysRevD.98.032011}. The off-axis cone method was also applied to substract the underlying event backgrounds, eg., beam remnants, in the analysis.\par

\section{Results}
\begin{figure*}
\centering
\begin{minipage}{.49\textwidth}
    \includegraphics[width=1\columnwidth]{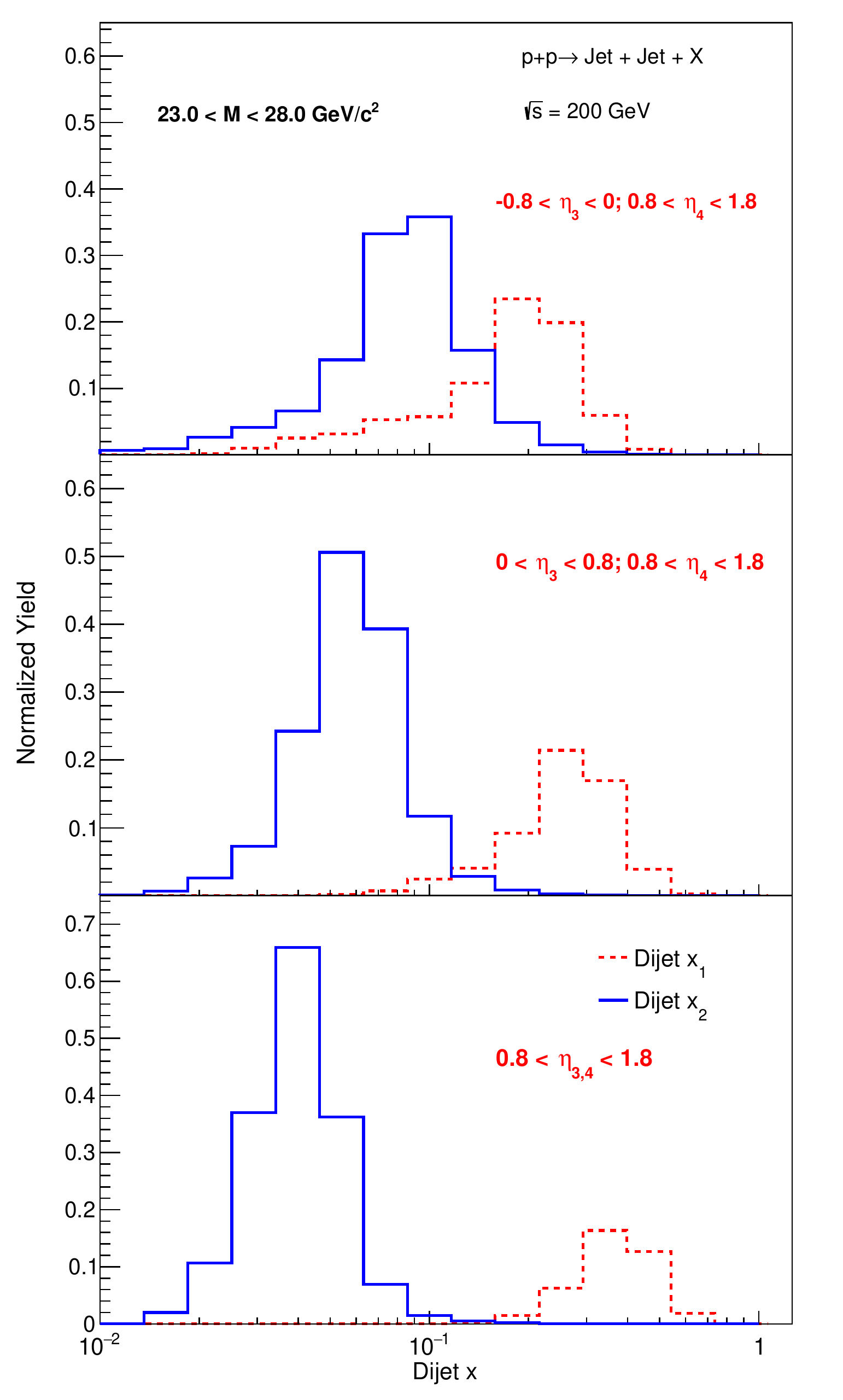}
    \caption{The distributions of the parton $x_1$ and $x_2$, which have been weighted by the partonic $\hat{a}_{LL}$, from \textsc{Pythia} simulations at $\sqrt{s}$ = 200 GeV for three different dijet topologies.}
    \label{fig:Dijet_Kinematics}
\end{minipage}
\hfill
\begin{minipage}{.49\textwidth}
  \includegraphics[width=1\columnwidth]{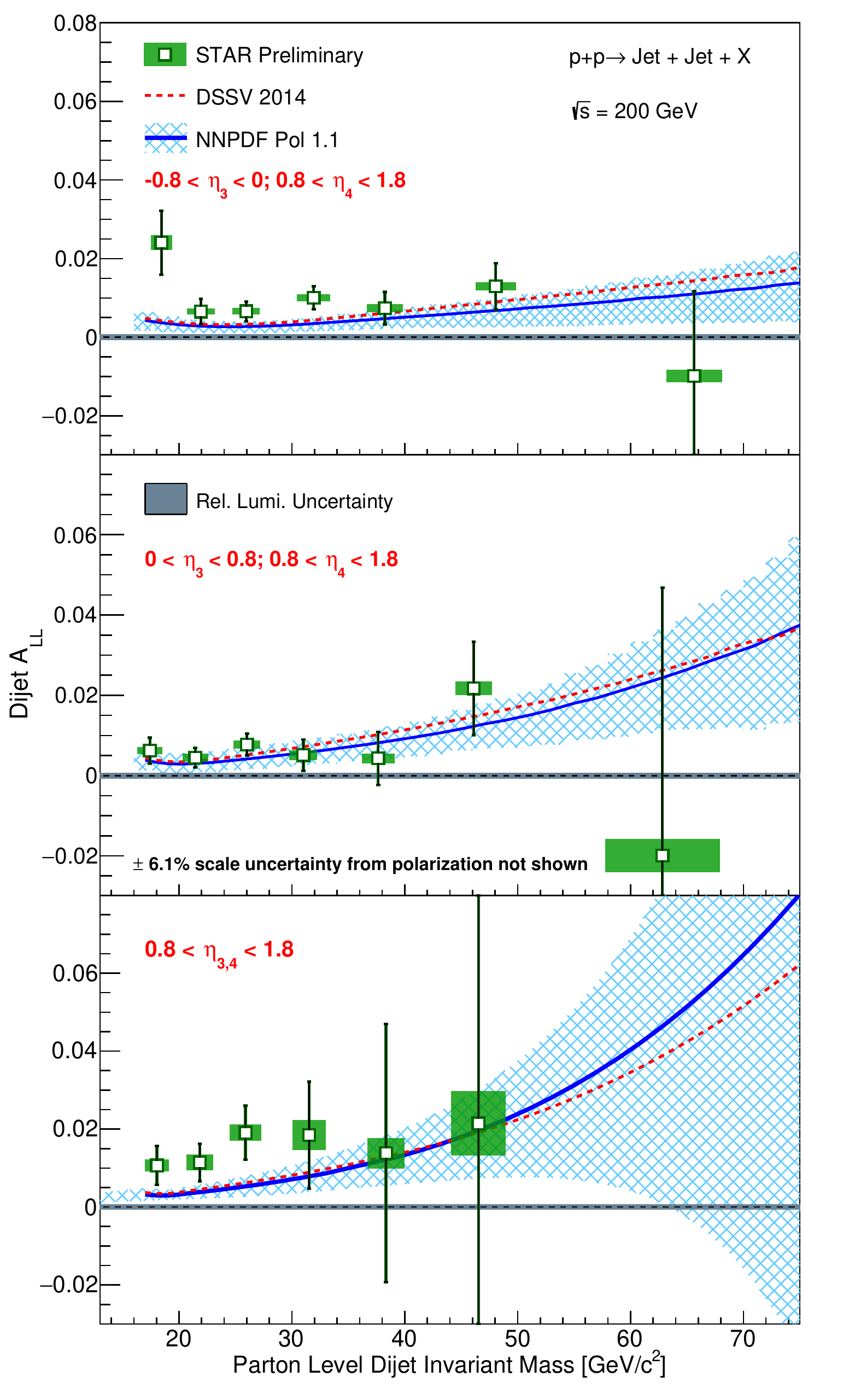}
    \caption{$A_{LL}$ as a function of parton-level dijet invariant mass for dijets with the East barrel-endcap (top), West barrel-endcap (middle) and endcap-endcap (bottom) event topologies.}
  \label{fig:Dijet_ALL_Run15_Run9}
\end{minipage}
\end{figure*}

As shown in Fig.~\ref{fig:Dijet_Kinematics}, dijets from more asymmetric collisions were chosen as the measurements extend into higher pseudorapidity. When both jets are in the Endcap ($0.8 < \eta_{\rm jet} < 1.8$), we are able to probe the hard scattering interaction between gluons and valence quarks with $x_{2} \sim 0.04$ and $x_{1} \sim 0.35$ at dijet mass bin $23.0 < M < 28.0~\mathrm{GeV}/c^{2}$. New preliminary results for longitudinal double spin asymmetries from dijet production are presented in Fig.~\ref{fig:Dijet_ALL_Run15_Run9}. These results are plotted as a function of dijet invariant mass, and are separated into three different dijet topologies based on the pseudorapidity of the jets. Two different theoretical expectations from DSSV2014~\cite{deFlorian:2014yva} and NNPDFpol1.1~\cite{Nocera:2014gqa} are also presented in Fig.~\ref{fig:Dijet_ALL_Run15_Run9}, both include the data from the STAR 2009 inclusive jet analysis~\cite{Adamczyk:2015}. Generally good agreements have been found for data and both theoretical expectations, and there is no clear preference of a polarized PDF global fit within the uncertainties.\par

\section{Conclusion}
We presented a new preliminary result of the longitudinal double spin asymmetry $A_{LL}$ for dijets production at intermediate pseudorapidity ($0.8 < \eta < 1.8$). The results are in good agreement with two theoretical predictions that support positive gluon polarization inside proton. Increasingly tight constraints have been placed on the gluon helicity distribution through the measurement of $A_{LL}$ over a wide range of kinematic regimes and different collision energies. These results, once included in the global analysis, will help to further constrain the value and shape of the gluon polarization inside the proton.\par

\end{document}